# Coupling Si$_3$N$_4$ waveguide to SOI waveguide using transformation optics

S. Hadi Badri*, M. M. Gilarlue

Department of Electrical Engineering, Sarab Branch, Islamic Azad University, Sarab, Iran

* sh.badri@iaut.ac.ir

## Abstract

Silicon nitride (Si$_3$N$_4$) planar waveguide platform combined with silicon-on-insulator (SOI) devices offer a whole new generation of system-on-chip applications. Therefore, efficient coupling of an Si$_3$N$_4$ waveguide to an SOI waveguide is essential. We present a coupler to interface a 1.8 µm-wide Si$_3$N$_4$ waveguide to a 0.5 µm-wide SOI waveguide based on the focusing property of the Luneburg lens. In order to match the refractive indices of the waveguides with the edges of the lens, one side of the lens is flattened with quasi-conformal transformation optics. The designed coupler is implemented by graded photonic crystals. The three-dimensional numerical simulations indicate that the 1.93 µm-long coupler has an average coupling loss of 0.13 dB in the C-band.

## Key words

## 1. Introduction

A broad class of photonic integrated devices based on the silicon nitride waveguide platform have been introduced due to its transparency over a wide wavelength range and compatibility with complementary metal-oxide-semiconductor (CMOS) foundry processes [1]. The refractive index contrast is lower in Si$_3$N$_4$ waveguides compared to SOI waveguides resulting in lower scattering losses associated with sidewall roughness, and therefore Si$_3$N$_4$ waveguides are more tolerant to fabrication imperfections. The operating range of SOI waveguides is limited by silicon (Si) bandgap (1.1 µm) and mid-infrared absorption of silica (3.7 µm). On the contrary, Si$_3$N$_4$ is transparent in the visible range making it a viable candidate for applications requiring shorter wavelengths [2]. Because of the larger bandgap of $Si_3N_4$ compared to Si, the two-photon absorption (TPA) and free-carrier absorption are absent in $Si_3N_4$ and, therefore, $Si_3N_4$ waveguides can operate at higher power levels [3]. Consequently, the Kerr nonlinearity of the $Si_3N_4$ has been exploited to implement frequency comb generation [4] and supercontinuum generation [5]. Different functions such as polarizers [6], grating couplers [7], multiple-wavelength oscillator [8], modulators [9], arrayed-waveguide gratings [10], resonators [11], continuous wave-pumped wavelength converters [12], and biosensors [13] have also been implemented in $Si_3N_4$ platform. On the other hand, SOI platform, due to its high index-contrast and compatibility with CMOS foundry processes, allows us to implement functions which are very difficult to implement in low index-contrast waveguide platforms, such as photonic crystals (PhC), grating couplers, and wavelength size cavities with a smaller footprint [14]. In order to enjoy the complementary features of $Si_3N_4$ and SOI platforms, efficient coupler with a small footprint should be designed. However, few reports have studied $Si_3N_4$ to SOI couplers. Interlayer couplers have been presented to couple an Si waveguide from one layer to an $Si_3N_4$ in another layer in multilayer platforms [15-18]. Optical couplers have also been designed to couple an $Si_3N_4$ waveguide to polymer [19], chalcogenide [20], and silicon-on-sapphire [21] waveguides.

Gradient index (GRIN) lenses such as Maxwell's fisheye [22, 23], Luneburg [24-26], and Eaton [27, 28] lenses have been employed to design various integrated photonic components. Luneburg lens focuses the parallel rays incident on its side to a point on the other side of the lens. The focusing property of the Luneburg lens has been utilized to design optical couplers [24-26]. In this paper, we exploit the focusing ability of the Luneburg lens to design a coupler interfacing an $Si_3N_4$ waveguide to an SOI waveguide. Because of the difference between the refractive indices of $Si_3N_4$ and Si, they cannot be coupled directly. To overcome this issue, we apply quasi-conformal transformation optics (QCTO) to flatten one side of the lens and, accordingly, increase the refractive index at this side of the lens. A 1.8 μm×0.22 μm (width×thickness) $Si_3N_4$ waveguide is coupled to a 0.5 μm×0.22 μm SOI waveguide through the flattened lens. The three-dimensional (3D) simulations reveal that the designed 1.93 μm-long coupler, implemented by graded photonic crystals (GPC), has an average coupling loss of 0.13 dB in the C-band.

## 2. Flattened Luneburg coupler

In this section, we design a coupler to interface an $Si_3N_4$ waveguide to an SOI waveguide based on the Luneburg lens and transformation optics [29, 30]. Typically, the width of waveguides with moderate refractive contrast such as $Si_3N_4$ waveguide is larger. The widths of the $Si_3N_4$ waveguides in references [3-12] are 500-11000 nm. On the other hand, the width of the SOI waveguides is typically about 500 nm. In this paper, we choose the widths of the $Si_3N_4$ and SOI waveguides as $W_{Si_3N_4}$ =1800 nm and $W_{SOI}$=500 nm, respectively, while the thickness of both waveguides is chosen as $H$=220 nm. The refractive index profile of the generalized Luneburg lens is described by [31]

$$n_{lens}(r) = n_{edge}\sqrt{1+f^2-(r/R_{lens})^2}/f \quad , \quad (0 \leq r \leq R_{lens}) \qquad (1)$$

where $n_{edge}$ is the refractive index of the lens at its edge, $r$ is the radial distance from the center, $R_{lens}$ is the radius of the lens, and $f$ determines the position of the focal point. For $f$ =1, the focal point lies on the edge of the lens while for $f >1$ and $f <1$ the focal point of the lens is located outside and inside of the lens, respectively. In this study, we use $f=1$ in our calculations and choose $R_{lens}$=1 μm and match the refractive index of the Si3N4 waveguide's core to the edge of the lens ($n_{edge}$=1.98). However, there is a refractive index mismatch at the interface of the lens's edge and the SOI waveguide. To alleviate this mismatch, we apply QCTO to flatten one side of the lens resulting in an increase in the refractive index at this side of the lens. Flattening of the Luneburg lens have been described in [32-36] so we do not repeat the procedure in here. As shown in Fig. 1, QCTO provides a method to transform a virtual domain (circular lens) to a physical domain (flattened lens) with minimum anisotropy. The Dirichlet and Neumann boundaries are shown as green and red boundaries in Fig. 1. In the virtual domain, Fig. 1(a), $\theta = 35°$ determines these boundaries. In the physical domain, Fig. 1(b), the domain's shape as well as the Dirichlet and Neumann boundaries are determined by $x_p = 0.88 \times R_{lens}$ and $y_p = -0.744 \times R_{lens}$. In order to reduce the footprint of the coupler and avoid the implementation of refractive indices lower than unity at the corners of the transformed lens, we truncate the lens as shown in Fig. 2(a). The lens is truncated in a shape of a parabolic taper where the widths of its sides correspond to the widths of the Si3N4 and SOI waveguides. The ray-tracing calculations indicate that the truncation of the lens has a limited effect on the performance of the lens as long as the incident rays are within the left and right edges of the lens.

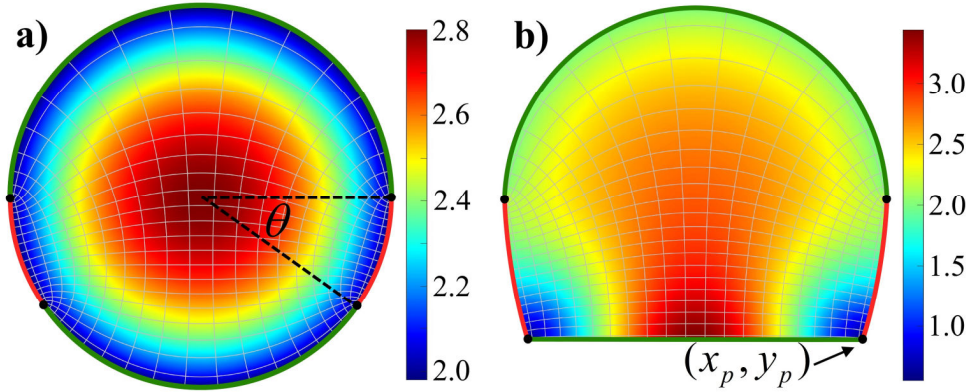

Fig. 1. The refractive indices and the orthogonal grids of the a) circular Luneburg lens (virtual domain) and b) flattened Luneburg lens (physical domain). The Dirichlet and Neumann boundaries are shown as green and red boundaries.

GRIN lenses have been implemented by multilayer structures [37-39], varying the guiding layer's thickness [40-42], and GPC [43, 44]. We implement the truncated lens with a GPC structure as shown in Fig. 2(b). First, we grid the lens into rectangular cells with side lengths of $a_x$=225 nm and $a_y$=250 nm. Then the average refractive index in each cell is calculated. Since the size of the cell is far smaller than the wavelength of the incident light (1550 nm), the medium can be regarded as a homogeneous medium with an effective refractive index calculated by effective medium theory (EMT) [44, 45]. Based on EMT calculations, an Si rod is placed at the center of a cell in an Si3N4 host. For the transverse electric (TE) mode, where the electric field is normal to the Si rods, the radius of the rod in the ij-th cell is calculated by [44]

$$r_{rod,ij} = a_{GPC} \sqrt{\frac{(\varepsilon_{eff,ij}^{TE} - \varepsilon_{host})(\varepsilon_{rod} + \varepsilon_{host})}{\pi(\varepsilon_{eff,ij}^{TE} + \varepsilon_{host})(\varepsilon_{rod} - \varepsilon_{host})}} \qquad (2)$$

where $\varepsilon_{eff}^{TE}$ is the effective permittivity of the cell for TE mode, $\varepsilon_{rod}$ and $\varepsilon_{host}$ are the permittivities of the rods (Si) and the host ($Si_3N_4$), respectively. The radius of the rod corresponding to the effective refractive index of a unit cell is calculated based on Eq. (2). In this calculation, a unit rectangular cell consisting of a silicon rod in the $Si_3N_4$ background with $a_x$=225 nm and $a_y$=250 nm is used. The radius of the rod with respect to the effective refractive index of the unit cell is illustrated in Fig. 3. As it is apparent in Fig. 2(b), some rods near the edges of the truncated lens are displaced to some extent in order to position them inside the lens. The radius of the smallest rod is 50 nm while the radius of the largest rod is 109 nm. It should be noted that the refractive index of the lens at its flattened edge ranges from 2.9 to 3.45 [Fig. 2(a)] while the refractive index of the SOI waveguide's core is 3.45. The numerical simulations indicate that the introduced mismatch has a negligible effect on the performance of the coupler since the average refractive index at the flattened edge is used in the design of the GPC structure. We can reduce the mismatch range at the cost of increasing the radius of the lens and, consequently, increasing the footprint of the designed coupler. The designed coupler, as well as the $Si_3N_4$ and SOI waveguides, are illustrated in Fig. 4. Silica is considered as the lower and upper cladding layers. The length of the flattened lens is about 1.75 μm. However, the effective length of the coupler, $L$=1.93 μm, is slightly longer than the lens since we filleted the corner of the structure where the SOI waveguide and the lens meet. The radius of the fillet is $r_{fillet}$=0.5 μm. A possible method for fabricating the proposed coupler is to use the reactive ion etching (RIE) [46] or electron-beam lithography combined with inductively coupled etching [47] to create the desired pattern of silicon rods on the $SiO_2$ substrate. Then silicon nitride can be deposited either by low pressure chemical vapor deposition (LPCVD) [10] or by plasma-enhanced chemical vapor deposition (PECVD) [48].

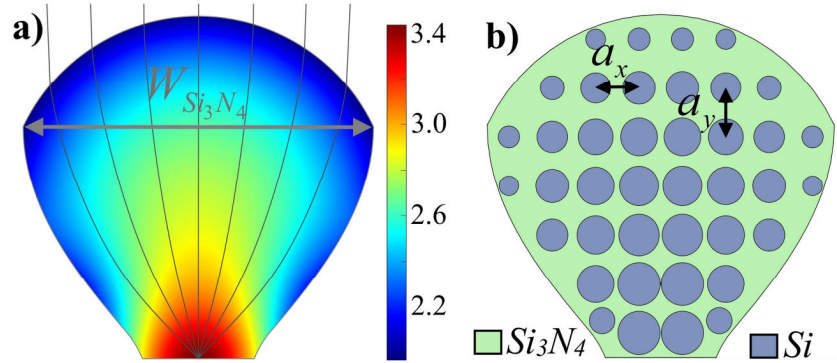

Fig. 2. a) The transformed lens [Fig. 1(a)] is truncated to reduce the footprint of the coupler. The refractive index of the truncated lens is also displayed. b) The truncated lens is implemented by a GPC structure consisting of Si rods in the $Si_3N_4$ background.

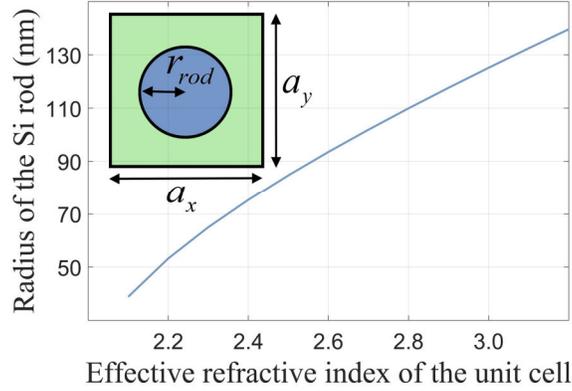

Fig. 3. Radius of the silicon rod versus the effective refractive index of the unit cell.

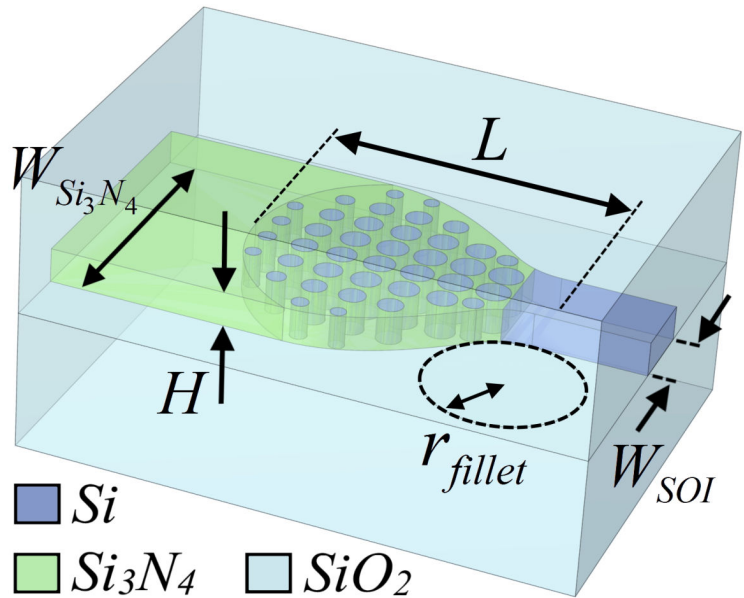

Fig. 4. The structure of the designed coupler interfacing an Si₃N₄ waveguide to an SOI waveguide.

## 3. Results and discussion

We employed Comsol Multiphysics ® for QCTO calculations, ray-tracing, and generating the electric field distribution figure. Since the 3D finite-difference time-domain (FDTD) calculations require less memory, Lumerical FDTD ® was utilized to calculate the scattering parameters. In FDTD simulations, the maximum mesh size in the lateral direction was 15 nm while in the vertical direction was 10 nm. The electric field distribution of TE mode light in the coupler at the wavelength of 1550 nm is displayed in Fig. 5. At this wavelength, the coupling loss of the coupler is about 0.13 dB while the return loss is about 21.5 dB. As illustrated in Fig. 5, the optical confinement in the Si₃N₄ waveguide is considerably lower than SOI waveguide, therefore, matching the electric fields of Si₃N₄ and SOI waveguides is challenging. The focusing property of the Luneburg lens facilitates the conversion from the $Si_3N_4$ waveguide mode to the SOI waveguide mode with a compact structure.

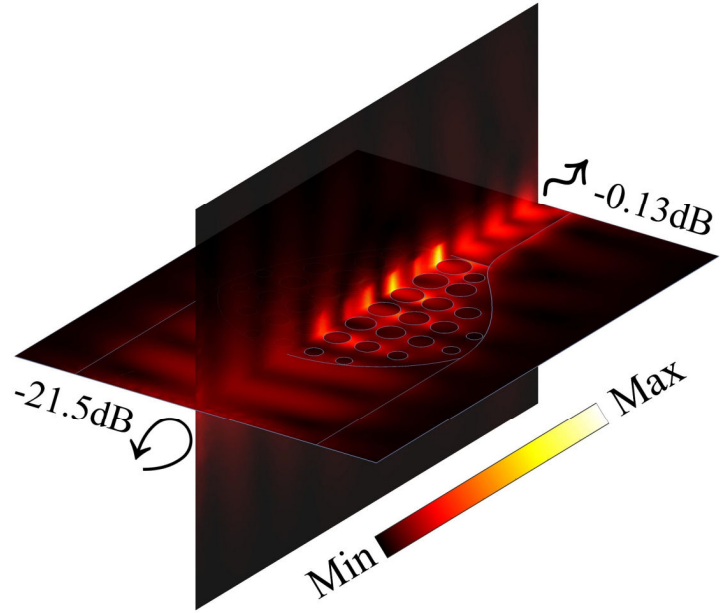

Fig. 5. Propagation of the TE mode light at 1550 nm from $Si_3N_4$ to SOI waveguide through the designed coupler.

The scattering parameters of the designed coupler are shown in Fig. 6. The optical signal is efficiently coupled from the $Si_3N_4$ waveguide to the SOI waveguide with an average coupling loss of about 0.13 dB in the C-band. The return loss of the coupler in the C-band is higher than 20 dB. Applying a finer grid in the GPC structure implementation ($a_{GPC}$<225 nm), slightly improves the coupling efficiency, however, this improvement is negligible. $a_{GPC}$ is the minimum of $a_x$ and $a_y$. On the other hand, as the $a_{GPC}$ increases, the coupling efficiency of the coupler degrades. For instance, for $a_{GPC}$=290 nm, the maximum coupling loss increases to 0.86 dB in the C-band.

Finally, we compare our coupler with previous studies. The introduced interlayer couplers in the multilayer platforms are typically longer than 20 µm [15-18]. The measured coupling loss in these interlayer couplers are 0.02-0.6 dB. The length of the adiabatic vertical tapering of the $Si_3N_4$ waveguide to the polymeric waveguide is 700 µm while the measured coupling loss is 0.14 dB [19]. A silicon inverse taper with a length of 150 µm couples a silicon-on-sapphire to silicon nitride waveguide [21]. The measured coupling loss at the wavelength of 1.5 µm is 4.8 dB while the simulations indicate that the coupling loss is 0.8 dB at the wavelength of 2 µm. Compared to these studies our proposed coupler, due to the focusing property of the Luneburg lens, is considerably shorter, i.e., 1.93 µm. Moreover, the designed structure has a coupling loss of 0.13 dB which is comparable with previous studies.

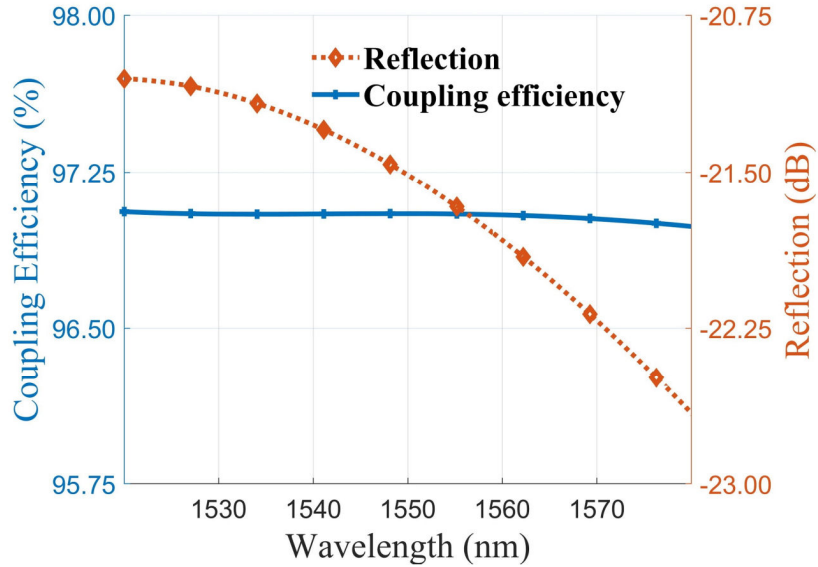

Fig. 6. The coupling efficiency and the reflection of the designed coupler.

## 4. Conclusion

A broad class of devices has been introduced by $Si_3N_4$ and SOI platforms. In order to benefit from the advantages of $Si_3N_4$ and SOI platforms simultaneously, efficient coupling between the $Si_3N_4$ and SOI waveguides is essential. We propose a coupler based on the Luneburg lens flattened by quasi-conformal transformation optics to increase the refractive index of the lens at its flattened edge. Therefore, the refractive index mismatch at the interfaces of the waveguides and the lens is reduced considerably. The designed 1.93 µm-long coupler efficiently couples the 1.8 µm-wide $Si_3N_4$ waveguide to the 0.5 µm-wide SOI waveguide. The 3D numerical simulations verify that the presented coupler has the average coupling efficiency of 97% in the C-band.

**Reference**

[1] D. J. Blumenthal, R. Heideman, D. Geuzebroek, A. Leinse, and C. Roeloffzen, "Silicon nitride in silicon photonics," *Proceedings of the IEEE,* vol. 106, no. 12, pp. 2209-2231, 2018.
[2] S. Romero-García, F. Merget, F. Zhong, H. Finkelstein, and J. Witzens, "Silicon nitride CMOS-compatible platform for integrated photonics applications at visible wavelengths," *Optics express,* vol. 21, no. 12, pp. 14036-14046, 2013.
[3] Y. Huang, Q. Zhao, L. Kamyab, A. Rostami, F. Capolino, and O. Boyraz, "Sub-micron silicon nitride waveguide fabrication using conventional optical lithography," *Optics Express,* vol. 23, no. 5, pp. 6780-6786, 2015.
[4] D. J. Moss, R. Morandotti, A. L. Gaeta, and M. Lipson, "New CMOS-compatible platforms based on silicon nitride and Hydex for nonlinear optics," *Nature photonics,* vol. 7, no. 8, p. 597, 2013.
[5] H. Zhao *et al.*, "Visible-to-near-infrared octave spanning supercontinuum generation in a silicon nitride waveguide," *Optics letters,* vol. 40, no. 10, pp. 2177-2180, 2015.

[6]     J. F. Bauters, M. Heck, D. Dai, J. Barton, D. Blumenthal, and J. Bowers, "Ultralow-Loss Planar Si3N4 Waveguide Polarizers," *IEEE Photonics Journal,* vol. 5, no. 1, pp. 6600207-6600207, 2012.

[7]     A. Z. Subramanian, S. Selvaraja, P. Verheyen, A. Dhakal, K. Komorowska, and R. Baets, "Near-infrared grating couplers for silicon nitride photonic wires," *IEEE Photonics Technology Letters,* vol. 24, no. 19, pp. 1700-1703, 2012.

[8]     J. S. Levy, A. Gondarenko, M. A. Foster, A. C. Turner-Foster, A. L. Gaeta, and M. Lipson, "CMOS-compatible multiple-wavelength oscillator for on-chip optical interconnects," *Nature photonics,* vol. 4, no. 1, p. 37, 2010.

[9]     K. Alexander *et al.*, "Nanophotonic Pockels modulators on a silicon nitride platform," *Nature communications,* vol. 9, no. 1, p. 3444, 2018.

[10]    D. Dai *et al.*, "Low-loss Si3N4 arrayed-waveguide grating (de) multiplexer using nano-core optical waveguides," *Optics express,* vol. 19, no. 15, pp. 14130-14136, 2011.

[11]    D. T. Spencer, J. F. Bauters, M. J. Heck, and J. E. Bowers, "Integrated waveguide coupled Si3N4 resonators in the ultrahigh-Q regime," *Optica,* vol. 1, no. 3, pp. 153-157, 2014.

[12]    C. J. Krückel, P. A. Andrekson, D. T. Spencer, J. F. Bauters, M. J. Heck, and J. E. Bowers, "Continuous wave-pumped wavelength conversion in low-loss silicon nitride waveguides," *Optics letters,* vol. 40, no. 6, pp. 875-878, 2015.

[13]    P. Muellner, E. Melnik, G. Koppitsch, J. Kraft, F. Schrank, and R. Hainberger, "CMOS-compatible Si3N4 waveguides for optical biosensing," *Procedia engineering,* vol. 120, pp. 578-581, 2015.

[14]    R. Baets *et al.*, "Silicon Photonics: silicon nitride versus silicon-on-insulator," in *Optical Fiber Communication Conference*, 2016: Optical Society of America, p. Th3J. 1.

[15]    Y. Huang, J. Song, X. Luo, T.-Y. Liow, and G.-Q. Lo, "CMOS compatible monolithic multi-layer Si3N4-on-SOI platform for low-loss high performance silicon photonics dense integration," *Optics express,* vol. 22, no. 18, pp. 21859-21865, 2014.

[16]    M. Sodagar, R. Pourabolghasem, A. A. Eftekhar, and A. Adibi, "High-efficiency and wideband interlayer grating couplers in multilayer Si/SiO2/SiN platform for 3D integration of optical functionalities," *Optics express,* vol. 22, no. 14, pp. 16767-16777, 2014.

[17]    A. H. Hosseinnia, A. H. Atabaki, A. A. Eftekhar, and A. Adibi, "High-quality silicon on silicon nitride integrated optical platform with an octave-spanning adiabatic interlayer coupler," *Optics express,* vol. 23, no. 23, pp. 30297-30307, 2015.

[18]    P. Xu, Y. Zhang, S. Zhang, Y. Chen, and S. Yu, "SiNx–Si interlayer coupler using a gradient index metamaterial," *Optics letters,* vol. 44, no. 5, pp. 1230-1233, 2019.

[19]    J. Mu *et al.*, "Low-loss, broadband and high fabrication tolerant vertically tapered optical couplers for monolithic integration of Si 3 N 4 and polymer waveguides," *Optics letters,* vol. 42, no. 19, pp. 3812-3815, 2017.

[20]    J.-E. Tremblay *et al.*, "Large bandwidth silicon nitride spot-size converter for efficient supercontinuum coupling to chalcogenide waveguide," in *2017 Conference on Lasers and Electro-Optics (CLEO)*, 2017: IEEE, pp. 1-2.

[21]    S. Zlatanovic, B. W. Offord, M. W. Owen, R. Shimabukuro, and E. Jacobs, "Mode-converting coupler for silicon-on-sapphire devices," in *Silicon Photonics X*, 2015, vol. 9367: International Society for Optics and Photonics, p. 93670W.

[22]    M. M. Gilarlue and S. H. Badri, "Photonic crystal waveguide crossing based on transformation optics," *Optics Communications,* vol. 450, pp. 308-315, 2019.

[23]    S. H. Badri and M. M. Gilarlue, "Maxwell's fisheye lens as efficient power coupler between dissimilar photonic crystal waveguides," *Optik,* vol. 185, pp. 566-570, 2019.

[24]    B. Arigong *et al.*, "Design of wide-angle broadband luneburg lens based optical couplers for plasmonic slot nano-waveguides," *Journal of Applied Physics,* vol. 114, no. 14, p. 144301, 2013.

[25]	S. H. Badri and M. M. Gilarlue, "Coupling between silicon waveguide and metal-dielectric-metal plasmonic waveguide with lens-funnel structure," *Plasmonics,* 2019, doi: 10.1007/s11468-019-01085-7.
[26]	L. H. Gabrielli and M. Lipson, "Integrated Luneburg lens via ultra-strong index gradient on silicon," *Optics express,* vol. 19, no. 21, pp. 20122-20127, 2011.
[27]	S. H. Badri and M. M. Gilarlue, "Low-index-contrast waveguide bend based on truncated Eaton lens implemented by graded photonic crystals," *JOSA B,* vol. 36, no. 5, pp. 1288-1293, 2019.
[28]	S. H. Badri, H. R. Saghai, and H. Soofi, "Polymer multimode waveguide bend based on multilayered Eaton lens," *Applied Optics,* vol. 58, no. 19, pp. 5219-5224, 2019.
[29]	M. McCall *et al.*, "Roadmap on transformation optics," *Journal of Optics,* vol. 20, no. 6, p. 063001, 2018.
[30]	H. Eskandari, M. S. Majedi, and O. Quevedo-Teruel, "Elliptical Generalized Maxwell Fish-Eye Lens using Conformal Mapping," *New Journal of Physics,* 2019.
[31]	J. A. Lock, "Scattering of an electromagnetic plane wave by a Luneburg lens. I. Ray theory," *JOSA A,* vol. 25, no. 12, pp. 2971-2979, 2008.
[32]	N. Kundtz and D. R. Smith, "Extreme-angle broadband metamaterial lens," *Nature materials,* vol. 9, no. 2, p. 129, 2010.
[33]	J. Hunt *et al.*, "Planar, flattened Luneburg lens at infrared wavelengths," *Optics express,* vol. 20, no. 2, pp. 1706-1713, 2012.
[34]	L. Wu, X. Tian, H. Ma, M. Yin, and D. Li, "Broadband flattened Luneburg lens with ultra-wide angle based on a liquid medium," *Applied Physics Letters,* vol. 102, no. 7, p. 074103, 2013.
[35]	H. F. Ma and T. J. Cui, "Three-dimensional broadband and broad-angle transformation-optics lens," *Nature communications,* vol. 1, p. 124, 2010.
[36]	S. H. Badri and M. M. Gilarlue, "Ultrashort waveguide tapers based on Luneburg lens," *Journal of Optics,* vol. 21, no. 12, p. 125802, 2019.
[37]	M. M. Gilarlue, J. Nourinia, C. Ghobadi, S. H. Badri, and H. R. Saghai, "Multilayered Maxwell's fisheye lens as waveguide crossing," *Optics Communications,* vol. 435, pp. 385-393, Mar 15 2019.
[38]	S. H. Badri, M. M. Gilarlue, H. Soofi, and H. R. Saghai, "3 × 3 slot waveguide crossing based on Maxwell's fisheye lens," *Optical Engineering,* vol. 58, no. 9, p. 097102, 2019.
[39]	T.-H. Loh, Q. Wang, K.-T. Ng, Y.-C. Lai, and S.-T. Ho, "CMOS compatible integration of Si/SiO 2 multilayer GRIN lens optical mode size converter to Si wire waveguide," *Optics express,* vol. 20, no. 14, pp. 14769-14778, 2012.
[40]	S. H. Badri, H. R. Saghai, and H. Soofi, "Polygonal Maxwell's fisheye lens via transformation optics as multimode waveguide crossing," *Journal of Optics,* vol. 21, no. 6, p. 065102, 2019.
[41]	S. H. Badri, H. R. Saghai, and H. Soofi, "Multimode waveguide crossing based on a square Maxwell's fisheye lens," *Applied Optics,* vol. 58, no. 17, pp. 4647-4653, 2019.
[42]	A. Di Falco, S. C. Kehr, and U. Leonhardt, "Luneburg lens in silicon photonics," *Optics express,* vol. 19, no. 6, pp. 5156-5162, 2011.
[43]	X.-H. Sun, Y.-L. Wu, W. Liu, Y. Hao, and L.-D. Jiang, "Luneburg lens composed of sunflower-type graded photonic crystals," *Optics Communications,* vol. 315, pp. 367-373, 2014.
[44]	M. M. Gilarlue, S. H. Badri, H. Rasooli Saghai, J. Nourinia, and C. Ghobadi, "Photonic crystal waveguide intersection design based on Maxwell's fish-eye lens," *Photonics and Nanostructures - Fundamentals and Applications,* vol. 31, pp. 154-159, 2018.
[45]	B. Vasić, G. Isić, R. Gajić, and K. Hingerl, "Controlling electromagnetic fields with graded photonic crystals in metamaterial regime," *Optics Express,* vol. 18, no. 19, pp. 20321-20333, 2010.
[46]	T.-M. Shih *et al.*, "Supercollimation in photonic crystals composed of silicon rods," *Applied Physics Letters,* vol. 93, no. 13, p. 131111, 2008.

[47]  F. Qin, Z.-M. Meng, X.-L. Zhong, Y. Liu, and Z.-Y. Li, "Fabrication of semiconductor-polymer compound nonlinear photonic crystal slab with highly uniform infiltration based on nano-imprint lithography technique," *Optics express,* vol. 20, no. 12, pp. 13091-13099, 2012.
[48]  L. Wang, W. Xie, D. Van Thourhout, Y. Zhang, H. Yu, and S. Wang, "Nonlinear silicon nitride waveguides based on a PECVD deposition platform," *Optics express,* vol. 26, no. 8, pp. 9645-9654, 2018.